\documentclass[fleqn,12pt,twoside]{article}
\usepackage{espcrc1}
\usepackage{graphicx}

\title{Hadron Physics and Confinement Physics in Lattice QCD}

\author{
H. Suganuma\address[titech]{Faculty of Science, Tokyo Institute of Technology, Tokyo 152-8552, Japan},
K. Amemiya$^{\rm a}$,
H. Ichie$^{\rm a}$,
N. Ishii$^{\rm a}$,
H. Matsufuru\address[RCNP]{RCNP, Osaka University, Mihogaoka 10-1, Ibaraki, Osaka 567-0047, Japan},
Y.~Nakajima$^{\rm b}$,
Y.~Nemoto\address[YITP]{Yukawa Institute for Theoretical Physics, Kyoto University, Kyoto 606-8502, Japan},
M.~Oka$^{\rm a}$
and 
T.T.~Takahashi$^{\rm b}$
}
       
\begin{document}

\maketitle

\begin{abstract} 
We are aiming to construct Quark Hadron Physics and Confinement Physics 
based on QCD. Using SU(3)$_c$ lattice QCD, we are 
investigating the three-quark potential at $T=0$ and $T \ne 0$, 
mass spectra of positive and negative-parity baryons 
in the octet and the decuplet representations of the SU(3) flavor, 
glueball properties at $T=0$ and $T \ne 0$.  
We study also Confinement Physics using lattice QCD. 
In the maximally abelian (MA) gauge, the off-diagonal gluon amplitude 
is strongly suppressed, 
and then the off-diagonal gluon phase shows strong randomness, 
which leads to  a large effective off-diagonal gluon mass, 
$M_{\rm off} \simeq 1.2 {\rm GeV}$.
Due to the large off-diagonal gluon mass in the MA gauge,  
infrared QCD is abelianized like nonabelian Higgs theories.
In the MA gauge, there appears a macroscopic network of the monopole 
world-line covering the whole system. From the monopole current, 
we extract the dual gluon field $B_\mu$, and examine the longitudinal 
magnetic screening. We obtain $m_B \simeq$ 0.5 GeV in the infrared region, 
which indicates the dual Higgs mechanism by monopole condensation. 
From infrared abelian dominance and infrared monopole condensation, 
low-energy QCD in the MA gauge is described with the 
dual Ginzburg-Landau (DGL) theory. 
\end{abstract} 

\maketitle

\section{Quark Hadron Physics from Lattice QCD}

Quantum chromodynamics (QCD) established 
as the fundamental theory of the strong interaction takes  
a simple form \cite{GS94,CONF2000}, 
\begin{equation}
{\cal L}_{\rm QCD}=-{1 \over 2} {\rm tr} G_{\mu\nu}G^{\mu\nu}
+\bar q (i \hspace{0.0cm}{\not \hspace{-0.1cm} D}-m_q) q, 
\end{equation}
however, it is still hard to understand 
nonperturbative phenomena,  
such as color confinement and dynamical chiral-symmetry breaking,   
due to the infrared strong-coupling feature \cite{GS94,CONF2000}. 
In this decade, lattice QCD Monte Carlo calculations have been developed 
and have been mainly applied to 
(i) {\it hadron spectroscopy} and 
(ii) {\it finite temperature QCD phase transition}, with a great success. 
However, lattice QCD is applicable also to  
(iii) {\it Quark Hadron Physics} and (iv) {\it Confinement Physics}, 
as a useful and reliable method.  

Our group is aiming to construct Quark Hadron Physics 
and Confinement Physics based on lattice QCD.
Our strategy is to adopt lattice QCD calculations to 
the relevant quantities pointed out in Quark Hadron Physics 
or Confinement Physics, in order to obtain the reliable 
physical picture based on QCD.

In relation to Quark Hadron Physics, 
we are investigating several important subjects in Quark Hadron Physics 
using SU(3)$_c$ lattice QCD at the quenched level as follows. 
\begin{enumerate}
\item
We numerically derive the static three-quark (3Q) potential 
$V_{\rm 3Q}$ at $T=0$, 
which is responsible to the baryon properties, 
and find that $V_{\rm 3Q}$ is well reproduced 
by the sum of a constant, the Coulomb term and the 
linear confinement term proportional to the total flux-tube length,  
with the accuracy better than a few \% [3-5]. 
\item
From the Polyakov-loop correction, we obtain the Q-$\bar {\rm Q}$ potential 
and the 3Q potential at $T \ne 0$, 
which characterize  the hadron structure at $T \ne 0$. 
\item 
We measure the lowest mass of the positive and the negative parity 
baryons in singlet, octet and decuplet representations 
of the SU(3) flavor, respectively \cite{NMNS01}.
The experimentally observed $\Lambda(1405)$ is much lighter 
than the corresponding baryon with the mass of 1.6GeV on the lattice, 
which may suggest the ${\bar K}N$ molecule picture for $\Lambda(1405)$. 
The octet-decuplet baryon mass splitting is also investigated.
\item
We are investigating the glueball properties both at $T=0$ and at $T \ne 0$.
\end{enumerate}
Here, for the accurate measurement of the 3Q potential at $T=0$ and
the lowest hadron mass in each channel, we adopt 
the {\it smearing technique} which reduces the excited-state contamination.
Furthermore, to get maximal information on the temporal correlation, 
we use {\it anisotropic lattices} where the temporal lattice spacing 
$a_t$ is finer than the spatial lattice spacing $a_s$ \cite{NMNS01}. 
In particular, the anisotropic lattice is quite useful for 
the finite temperature QCD, because of the limitation of 
the temporal distance \cite{MMSU01}. 

Since these studies are introduced in [5-7],   
we mainly present the recent progress of our studies on 
Confinement Physics based on SU(2) lattice QCD at the quenched level. 

\section{Confinement Physics from Lattice QCD}

To understand the confinement mechanism is 
one of the most difficult problems remaining in the particle physics.
As the Regge trajectories and lattice QCD calculations indicate, 
quark confinement is 
characterized by {\it one-dimensional squeezing} of the 
color-electric flux and 
the {\it string tension} $\sigma \simeq 1{\rm GeV/fm}$, 
which is the key quantity of confinement.
On the confinement mechanism, 
Nambu first proposed the {\it dual superconductor theory} for quark 
confinement \cite{N74}, based on the electro-magnetic duality in 1974.  
In this theory, there occurs the one-dimensional squeezing 
of the color-electric flux  
by the {\it dual Meissner effect} due to condensation of 
bosonic color-magnetic monopoles. 
However, there are {\it two large gaps} between QCD and the 
dual superconductor theory [2,9-11]
\begin{enumerate} 
\item 
The dual superconductor theory is based on the {\it abelian gauge theory} 
subject to the Maxwell-type equations, where electro-magnetic duality is 
manifest, while QCD is a nonabelian gauge theory.   
\item
The dual superconductor theory requires color-magnetic monopole 
condensation as the key concept, while QCD does not have 
color magnetic monopoles as the elementary degrees of freedom.
\end{enumerate}
These gaps may be simultaneously fulfilled by taking 
{\it MA gauge fixing,} which reduces QCD to an abelian gauge theory 
including color-magnetic monopoles.

In Euclidean QCD, the maximally abelian (MA) gauge is 
defined so as to minimize the total amount of the 
off-diagonal gluons [2,9-11]
\begin{equation}
R_{\rm off} [A_\mu ( \cdot )] \equiv \int d^4x \ {\rm tr}
\left\{ 
[\hat D_\mu ,\vec H][\hat D_\mu ,\vec H]^\dagger 
\right\} 
={e^2 \over 2} \int d^4x \sum_\alpha |A_\mu ^\alpha (x)|^2 
\end{equation}
by the SU($N_c$) gauge transformation.
Here, we have used the Cartan decomposition, 
$A_\mu (x)=\vec A_\mu (x) \cdot \vec H 
+\sum_\alpha A_\mu^\alpha (x)E^\alpha $. 
Since the ${\rm SU}(N_c)$ covariant derivative operator 
$\hat D_\mu \equiv \hat \partial_\mu+ieA_\mu $ obeys the 
adjoint gauge transformation, the local form of 
the MA gauge condition is easily derived as 
$[\vec H, [\hat D_\mu , [\hat D_\mu , \vec H]]]=0$. 
In the MA gauge, the gauge symmetry 
$G \equiv {\rm SU}(N_c)_{\rm local}$ 
is reduced into $H \equiv {\rm U(1)}_{\rm local}^{N_c-1} 
\times {\rm Weyl}^{\rm global}_{N_c}$, 
where the global Weyl symmetry is a subgroup of ${\rm SU}(N_c)$ 
relating the permutation of $N_c$ bases in the fundamental 
representation. 
In the MA gauge, off-diagonal gluons behave as charged matter fields like 
$W_\mu^{\pm}$ in the Standard Model, and provide the 
color-electric current in terms of the residual abelian gauge symmetry. 
In addition, 
color-magnetic monopoles appear as topological objects reflecting 
the nontrivial homotopy group [2,9-14]
\begin{equation}
\Pi_2({\rm SU}(N_c)/{\rm U(1)}^{N_c-1})=\Pi_1({\rm U(1)}^{N_c-1})
={\bf Z}^{N_c-1}_\infty 
\end{equation}
in a similar manner to similarly in the GUT monopole.
Here, the global Weyl symmetry and color-magnetic monopoles are 
relics of nonabelian nature of QCD. 

In this way, in the MA gauge, QCD is reduced into an abelian gauge theory 
including color-magnetic monopoles, 
which is expected to provide a theoretical basis of the 
dual superconductor theory for quark confinement.
Furthermore, recent lattice QCD studies show remarkable features of 
{\it abelian dominance} and {\it monopole dominance} for 
nonperturbative QCD (NP-QCD) in the MA gauge.
\begin{enumerate}
\item
Without gauge fixing, all the gluon components equally contribute to 
NP-QCD, and it is difficult to extract relevant degrees of freedom for NP-QCD. 
\item
In the MA gauge, QCD is reduced into an abelian gauge theory 
including the electric current $j_\mu $ and the magnetic current $k_\mu $, 
which forms a  
{\it global network of the monopole world-line covering the whole system.}
(See Fig.3(a).) 
In the MA gauge, lattice QCD shows {\it abelian dominance} 
for NP-QCD (confinement \cite{IS9900}, gluon propagators \cite{AS99}, 
chiral symmetry breaking \cite{M95}): only the diagonal gluon 
is relevant for NP-QCD, while off-diagonal gluons do not 
contribute to NP-QCD. 
\item
By the Hodge decomposition, the diagonal gluon is decomposed into  
the ``photon part'' ($j_\mu \ne 0, k_\mu=0$) and the 
``monopole part'' ($k_\mu \ne 0$, $j_\mu=0$), 
corresponding to the separation of $j_\mu$ and $k_\mu$. 
In the MA gauge, lattice QCD shows 
{\it monopole dominance} [17-19] for NP-QCD: 
the monopole part leads to NP-QCD, 
while the photon part  seems trivial like QED 
and does not contribute to NP-QCD. 
For example, on the Q-$\bar{\rm  Q}$ potential, 
the purely linear potential appears in the monopole part, 
while the Coulomb potential appears in the photon part 
like QED \cite{P97}.
\end{enumerate}
Thus, in the MA gauge, QCD is reduced into an abelian gauge theory 
with color-magnetic monopoles, {\it keeping essence of 
infrared nonperturbative features}, and 
the {\it relevant collective mode for NP-QCD} 
can be extracted as the color-magnetic monopole.

\subsection{Strong Randomness of Off-diagonal Gluon Phase, Abelian Dominance 
and Large Mass of Off-diagonal Gluons in MA Gauge}

To find out essence of the MA gauge,  
we study the off-diagonal gluon field 
$A_\mu^\pm \equiv \frac1{\sqrt{2}}(A_\mu^1 \pm i A_\mu^2)$ 
in the MA gauge in SU(2) lattice QCD. 
There are two remarkable features in the off-diagonal gluon field 
$A_\mu^\pm(x)=e^{\pm i \chi_\mu(x)}|A_\mu^\pm(x)|$ 
in the MA gauge [2,9-11]. 
\begin{enumerate}
\item
The off-diagonal gluon amplitude $|A_\mu^{\pm}(x)|$ 
is strongly suppressed by SU($N_c$) gauge transformation in the MA gauge.
\item
The off-diagonal gluon phase $\chi_\mu(x)$ 
tends to be random, because $\chi_\mu(x)$ is not 
constrained by MA gauge fixing at all, 
and only the constraint from the QCD action is weak 
due to a small accompanying factor $|A_\mu^\pm|$.
\end{enumerate}
\begin{figure}
\begin{center}
\includegraphics[scale=0.50]{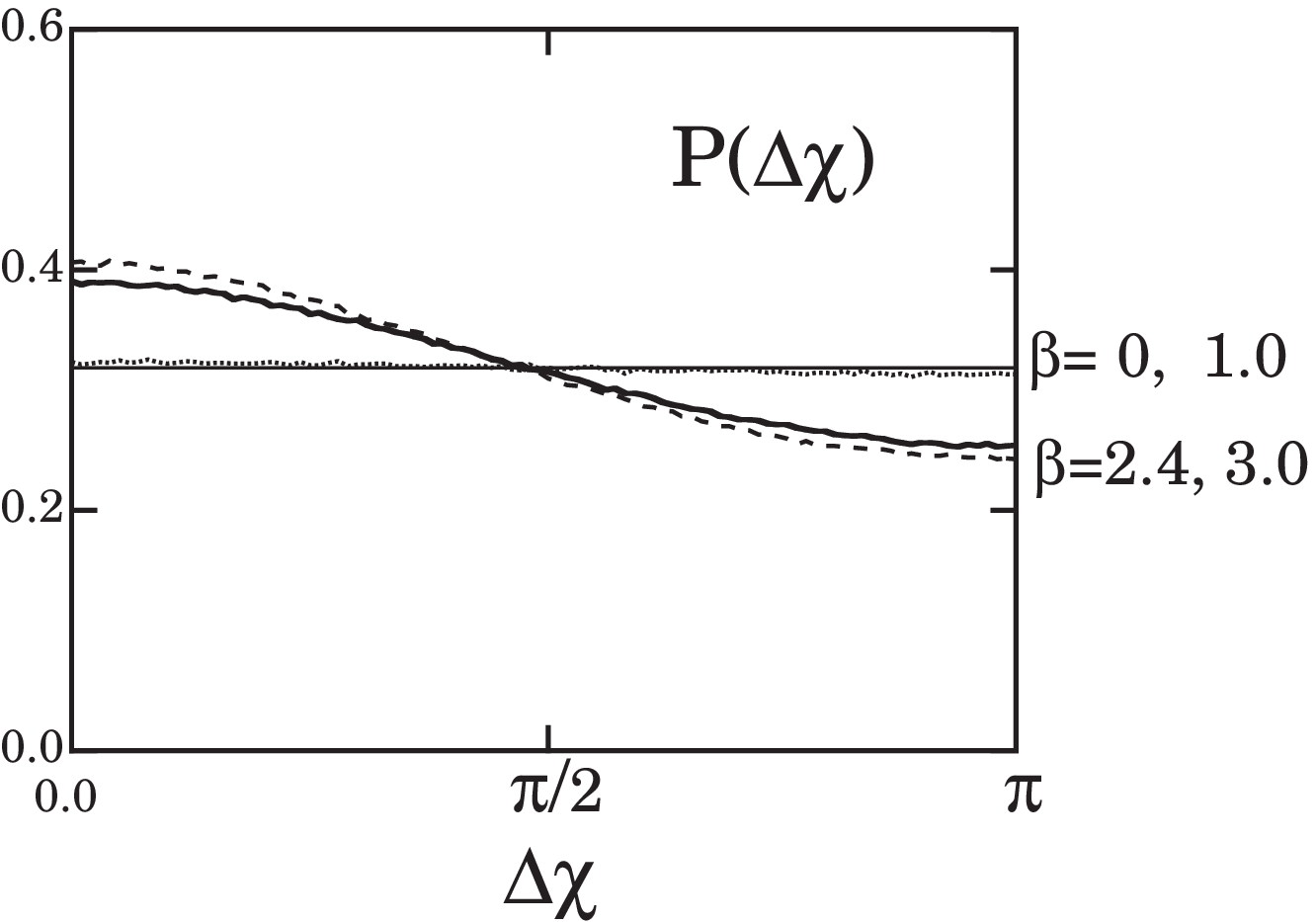}
\includegraphics[scale=0.40]{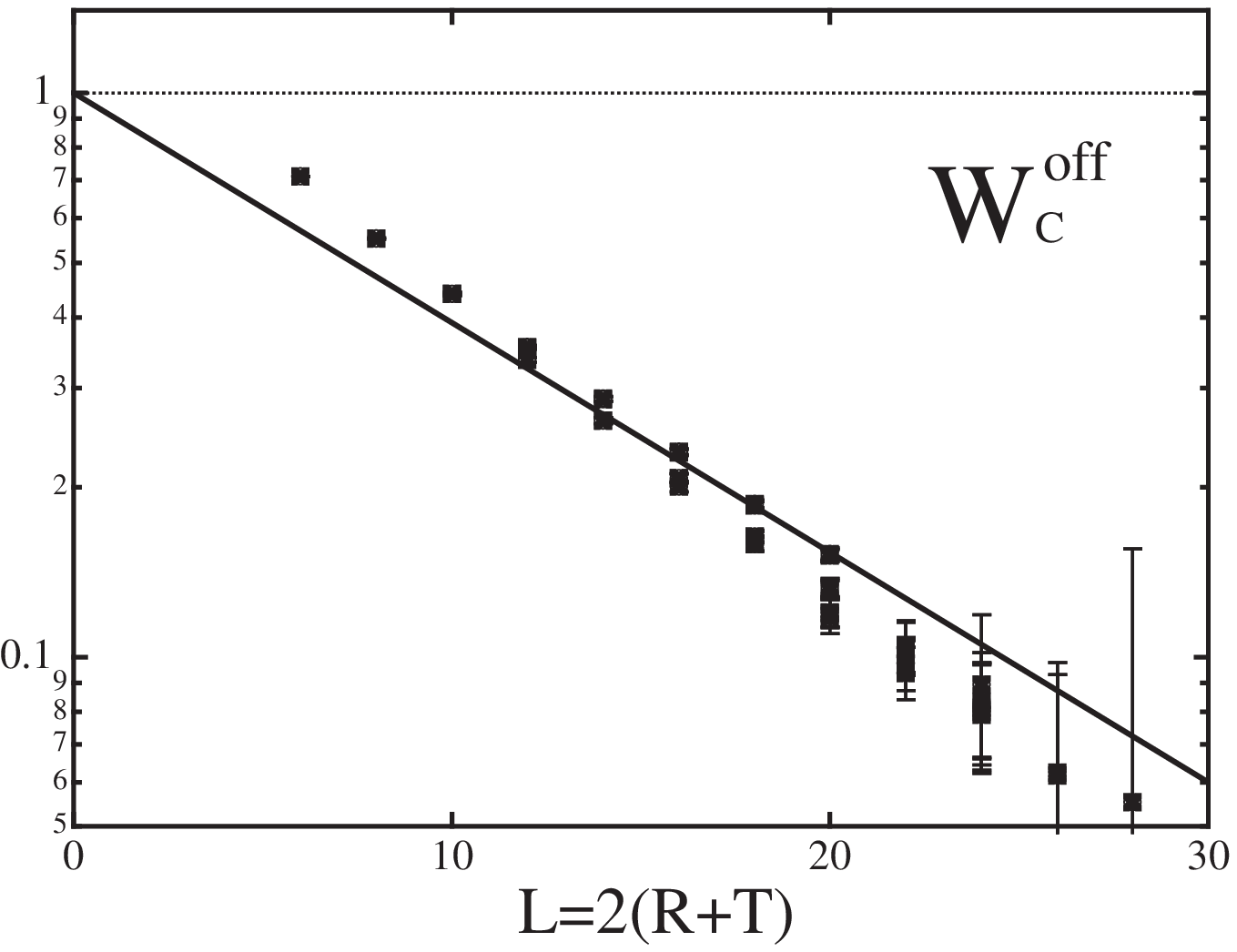}
\caption{
(a) The probability distribution $P(\Delta \chi)$ of the difference 
$\Delta \chi \equiv |\chi_\mu(s)-\chi_\mu(s+\hat \nu)| ({\rm mod } \pi)$ 
in the MA gauge plus U(1)$_3$ Landau gauge at $\beta$=0 
($a=\infty$, thin line), 
$\beta$=1.0 
($a \simeq$ 0.57fm, dotted curve), 
$\beta$=2.4 ($a \simeq$ 0.127fm, solid curve), 
$\beta$=3.0 ($a \simeq$ 0.04fm, dashed curve).
(b) The off-diagonal gluon contribution to the Wilson loop 
$W_C^{\rm off}$
v.s. the perimeter length $L \equiv 2(R+T)$
in the MA gauge on $16^4$ lattice with $\beta$=2.4.
The thick line denotes the theoretical estimation in Eq.(4).}
\end{center}
\end{figure}
Now, we consider the difference 
$\Delta \chi \equiv |\chi_\mu(s)-\chi_\mu(s+\hat \nu)| ({\rm mod } \pi)$ 
in the MA gauge. 
If the off-diagonal gluon phase  $\chi_\mu(x)$ is a continuum variable,  
as the lattice spacing $a$ goes to 0, 
$\Delta \chi \simeq a |\partial_\nu \chi_\mu|$ must 
go to zero, and hence $P(\Delta \chi)$ approaches to 
the $\delta$-function like $\delta(\Delta \chi)$. 
However, as shown in Fig.1(a), $P(\Delta \chi)$ is 
almost $a$-independent and almost flat. These features
indicate the {\it strong randomness of the off-diagonal gluon phase} 
$\chi_\mu(x)$ in the MA gauge. 
Then, $\chi_\mu(x)$ behaves as a random angle variable 
in the MA gauge.\footnote{
Near the monopole, a large amplitude of $|A^\pm_\mu(s)|$ remains 
even in the MA gauge, 
and $\chi_\mu(s)$ is constrained so as to reduce the QCD action. 
Hence, $\chi_\mu(s)$ cannot be regarded as a random 
variable near monopoles.}

Within the random-variable approximation 
for the off-diagonal gluon phase $\chi_\mu(s)$ in the MA gauge, 
we analytically prove abelian dominance of the string tension [2,9-11] 
from the analysis of the SU(2) Wilson loop $\langle W_C[A_\mu^a]\rangle$, 
the abelian Wilson loop 
$\langle W_C[A_\mu^\pm \equiv 0, A_\mu^3]\rangle_{\rm MA}$ and 
the off-diagonal gluon contribution to the Wilson loop 
$W_C^{\rm off} \equiv \langle W_C[A_\mu^a]\rangle/
\langle W_C[A_\mu^\pm \equiv 0, A_\mu^3]\rangle_{\rm MA}$. 
The point is the cancellation of 
the off-diagonal gluon contribution due to the random phase as   
$\langle e^{i\chi_\mu(s)} \rangle_{\rm MA} 
\simeq \frac{1}{2\pi} \int_0^{2\pi} d\chi_\mu(s) e^{i\chi_\mu(s)}=0$. 

Near the continuum limit $a \simeq 0$, we find 
a relation between the {\it macroscopic} quantity $W_C^{\rm off}$ 
and the {\it microscopic} quantity of the off-diagonal gluon amplitude 
$\langle |eA_\mu^\pm|^2 \rangle_{\rm MA}$ as 
\begin{equation}
W_C^{\rm off}\equiv 
\langle W_C[A_\mu^a]\rangle 
/\langle W_C[A_\mu^\pm \equiv0, A_\mu^3]\rangle_{\rm MA}
\simeq \exp\{L_{\rm phys} a \langle |eA_\mu^\pm|^2 \rangle_{\rm MA}/4\} 
\end{equation}
with the perimeter $L_{\rm phys} \equiv L a$ of the Wilson loop.
This relation is checked in lattice QCD as shown in Fig.1(b). 
In fact, the off-diagonal gluon contribution $W_C^{\rm off}$ obeys the 
{\it perimeter law},\footnote{
The off-diagonal contribution $W_C^{\rm off}$ 
becomes trivial as $W_C^{\rm off} \rightarrow 1$ 
in the continuum limit $a \rightarrow 0$.
}
and then {\it perfect abelian dominance for the string tension}, 
$\sigma_{\rm SU(2)}=\sigma_{\rm Abel}$, is derived 
by regarding off-diagonal gluon phases to be random in the MA gauge.

As another remarkable fact, 
{\it strong randomness of off-diagonal gluon phases leads to 
rapid reduction of off-diagonal gluon correlations.} 
In fact, if $\chi_\mu(x)$ is a complete random phase, 
Euclidean off-diagonal gluon propagators exhibit the 
$\delta$-functional reduction as 
\begin{equation}
\langle A_\mu^+(x) A_\nu^-(y) \rangle_{\rm MA} =
\langle |A_\mu^+(x)||A_\nu^-(y)|e^{i\{\chi_\mu(x)-\chi_\nu(y)\}} 
\rangle_{\rm MA} \nonumber =
\langle |A_\mu^\pm(x)|^2 \rangle_{\rm MA} \delta_{\mu\nu}\delta^4(x-y), 
\end{equation}
which means the infinitely large mass of off-diagonal gluons.
Of course, the real off-diagonal gluon phases are not complete 
but approximate random phases. 
Then, the off-diagonal gluon mass would be large but finite. 
Thus, {\it strong randomness of off-diagonal gluon phases 
is expected to provide a large effective mass of off-diagonal gluons.} 

\subsection{Large Mass Generation of Off-diagonal Gluons in MA Gauge : 
Essence of Infrared Abelianization of QCD}

We quantitatively study the Euclidean gluon propagator 
$G_{\mu \nu }^{ab} (x-y) \equiv \langle A_\mu ^a(x)A_\nu ^b(y)\rangle$ 
($a,b =1,2,3$) and the off-diagonal gluon mass $M_{\rm off}$ 
in the MA gauge, using SU(2) lattice QCD with $2.2 \le \beta \le 2.4$ 
and various sizes ($12^3 \times 24$, $16^4$, $20^4$). 
As for the residual U(1)$_3$ gauge symmetry, 
we take U(1)$_3$ Landau gauge, 
to extract most continuous gluon configuration under 
the MA gauge constraint, for the comparison with the continuum theory.
The continuum gluon field $A_\mu^a(x)$ is derived from 
the link variable as 
$U_\mu(s)={\rm exp}\{iaeA_\mu^a(s) \frac{\tau^a}{2}\}$.
We show in Fig.2(a) the scalar-type gluon propagators
$G_{\mu \mu}^3(r)$ and 
$G_{\mu\mu}^{+-}(r) \equiv \langle A_\mu^{+}(x)A_\mu^{-}(y)\rangle
= \frac12 \{G_{\mu\mu}^1(r)+G_{\mu\mu}^2(r)\}$, 
which depend only on the four-dimensional Euclidean 
distance $r \equiv \sqrt{(x_\mu- y_\mu)^2}$.
We find {\it infrared abelian dominance for the gluon propagator 
in the MA gauge}: only the abelian gluon $A_\mu^3(x)$ propagates over 
the long distance and can influence 
the infrared physics \cite{SAIT00,SITA98,AS99}.

\begin{figure}
\begin{center}
\includegraphics[scale=0.34]{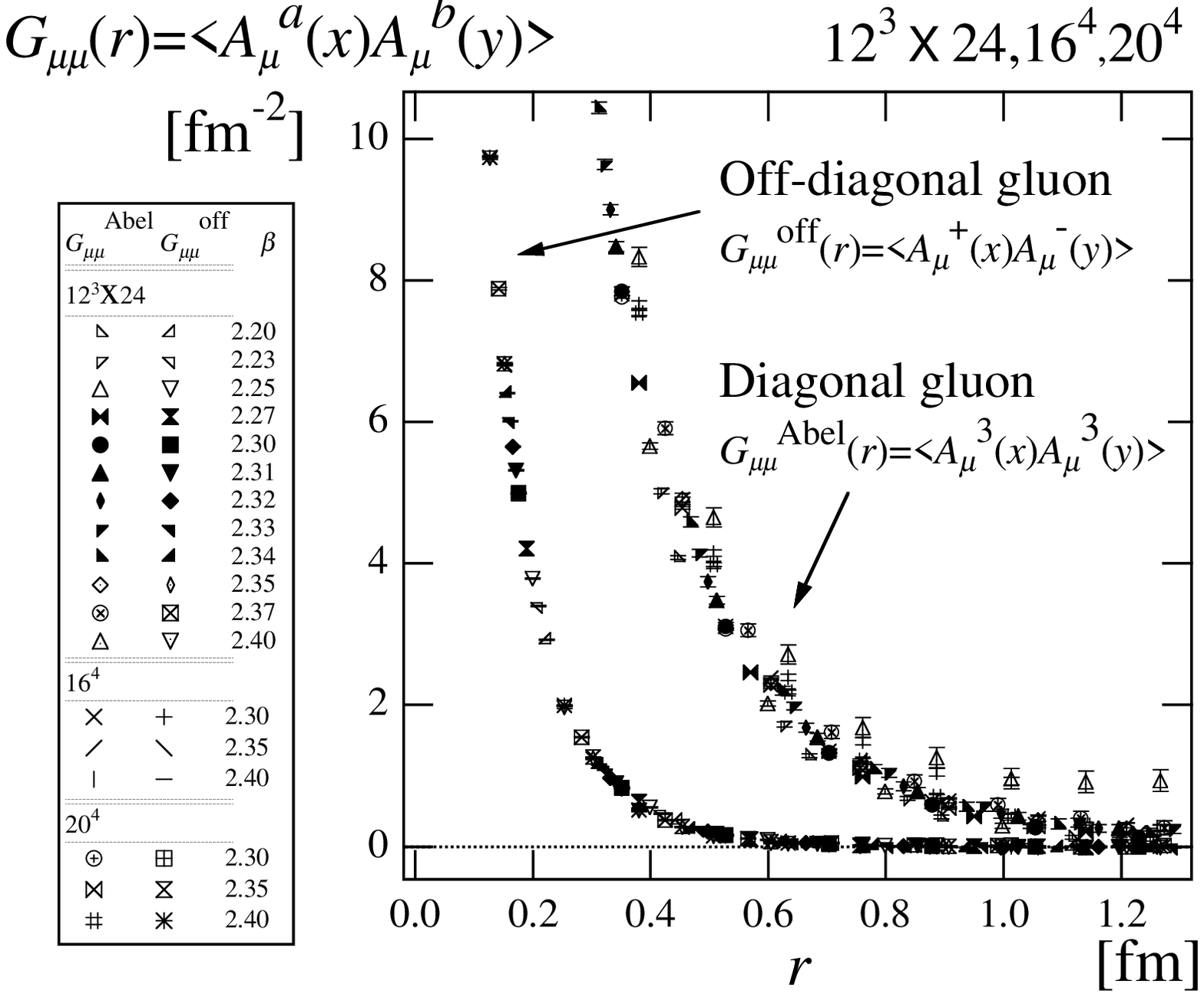}
\includegraphics[scale=0.34]{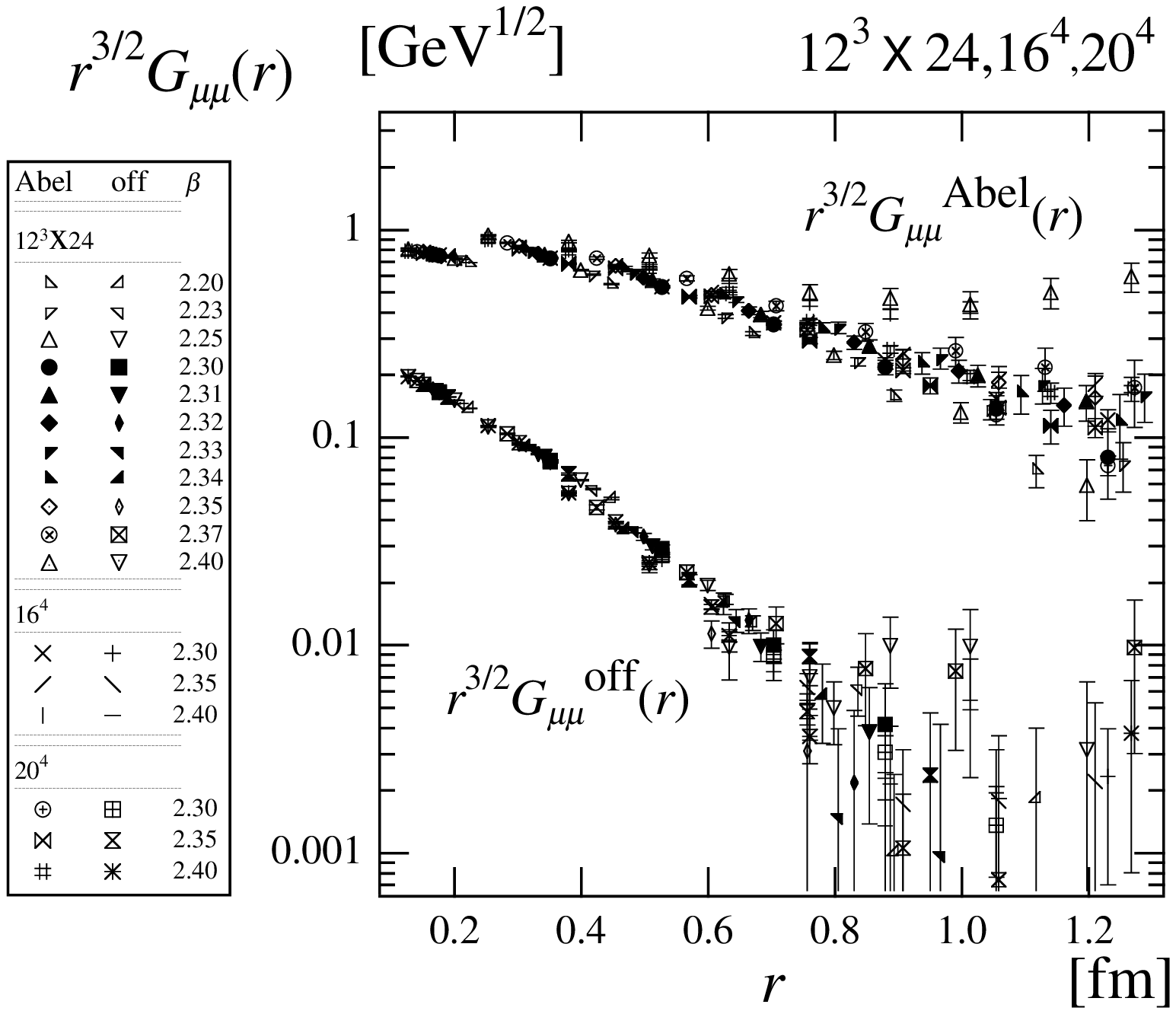}
\includegraphics[scale=0.4]{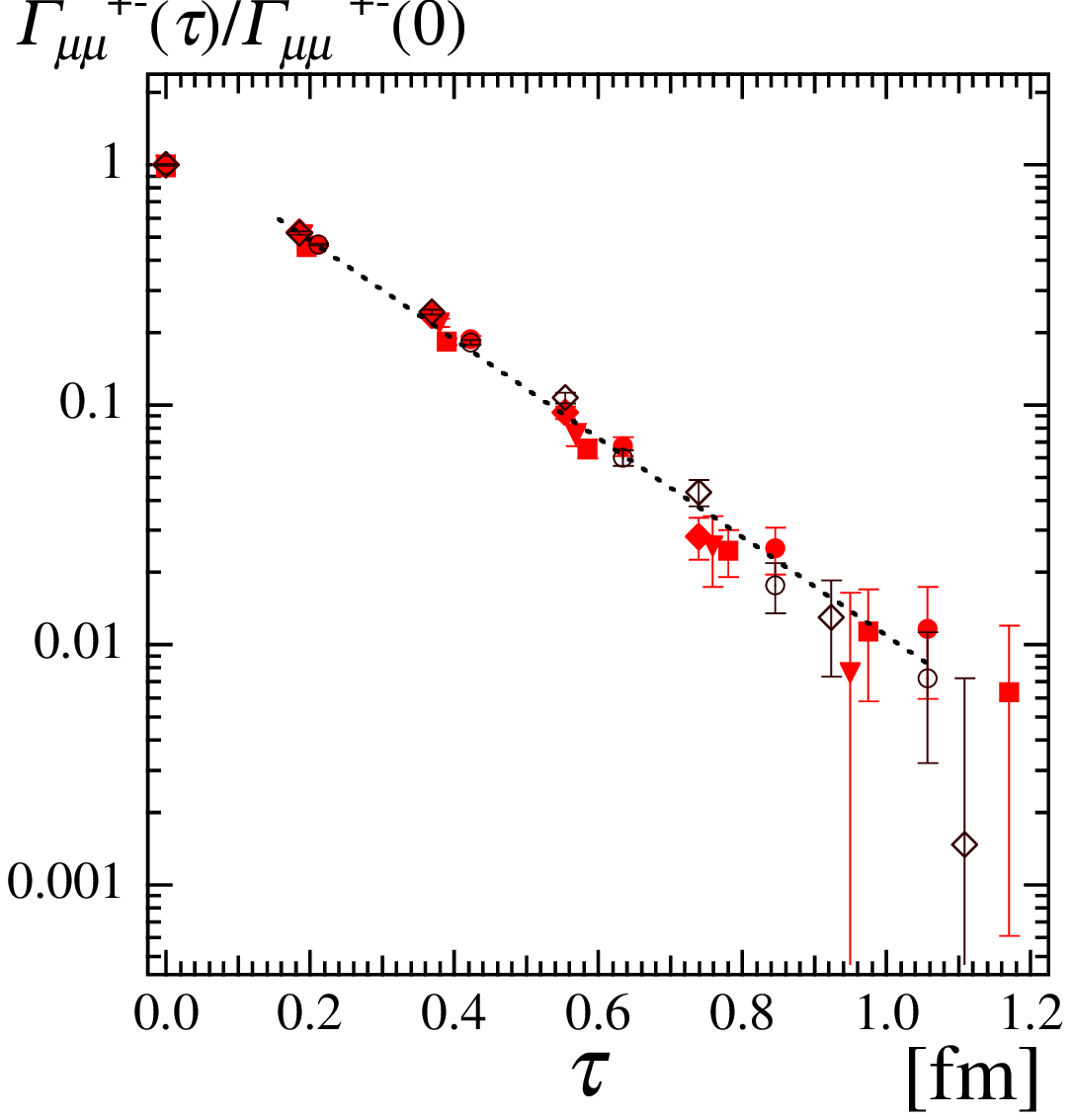}
\caption{ (a) The scalar-type gluon propagator 
$G_{\mu \mu }^a(r)$ as the function of the four-dimensional 
distance $r$ in the MA gauge in SU(2) lattice QCD with 
$2.2 \le \beta \le 2.4$ with various sizes 
($12^3 \times 24$, $16^4$, $20^4$). 
(b) The logarithmic plot of $r^{3/2} G_{\mu \mu}^a(r)$ v.s. $r$. 
The off-diagonal gluon propagator behaves as 
the Yukawa-type function, 
$G_{\mu \mu } \sim {\exp(-M_{\rm off}r) \over r^{3/2}}$. 
(c) The logarithmic plot of the temporal correlation 
$
\Gamma_{\mu\mu}^{+-}(\tau) \equiv 
\langle O_\mu^+(\tau)O_\mu^-(0) \rangle 
$
as the function of the temporal distance $\tau$ 
in SU(2) lattice QCD with $2.3 \le \beta \le 2.35$ 
with $16^3 \times 32$ and $12^3 \times 24$. 
From the slope in (b) and (c), 
the effective mass of the off-diagonal gluon $A_\mu^\pm$ 
is estimated as $M_{\rm off}\simeq 1.2 {\rm GeV}$.} 
\end{center}
\end{figure}

Since the four-dimensional Euclidean propagator of the 
massive vector boson with the mass $M$ takes a 
Yukawa-type asymptotic form as 
\begin{equation}
G_{\mu\mu}(r) = \frac3{4\pi^2} \frac{M}{r} K_1(Mr)+\frac1{M^2}\delta^4(x-y)
\simeq \frac{3M^{1/2}}{2(2\pi)^{3/2}}\frac{e^{-Mr}}{r^{3/2}},
\end{equation}
the infrared effective mass $M_{\rm off}$ of the off-diagonal gluon 
$A_\mu^{\pm}(x)$ can be extracted from the slope in the logarithmic plot of 
$r^{3/2} G_{\mu\mu}^{+-}(r)\sim \exp (-M_{\rm off}r)$ in Fig.2(b). 
From the slope analysis of the lattice QCD data with $r \ge 0.2 {\rm fm}$,  
we obtain the off-diagonal gluon mass as 
$M_{\rm off} \simeq 1.2~{\rm GeV}$ in the MA gauge.

We carry out also the {\it mass measurement of 
off-diagonal gluons from the temporal correlation of 
the zero-momentum projected operator} $O_\mu^\pm(\tau)$,  
\begin{equation}
\Gamma_{\mu\mu}^{+-}(\tau) \equiv 
\langle O_\mu^+(\tau)O_\mu^-(0) \rangle, 
\quad 
O_\mu^\pm(\tau) \equiv \int d{\bf x} \ A_\mu^\pm({\bf x},\tau), 
\end{equation}
in the MA gauge plus ${\rm U(1)}_3$ Landau gauge.
in SU(2) lattice QCD with $2.3 \le \beta \le 2.35$ with 
$16^3\times 32$ and $12^3\times 24$.
Again, we find the off-diagonal gluon mass $M_{\rm off} \simeq 1.2 {\rm GeV}$ 
in the MA gauge from the slope of the logarithmic plot of 
$\Gamma_{\mu\mu}^{+-}(\tau)$ in Fig.2(c) \cite{CONF2000,SAIT00}. 

Thus, the {\it off-diagonal gluon $A_\mu^\pm$ acquires a large effective mass  
$M_{\rm off} \simeq 1.2 {\rm GeV}$ in the MA gauge}, which is 
{\it essence of infrared abelian dominance} \cite{SAIT00,SITA98,AS99}.
In the MA gauge, 
due to the large effective mass $M_{\rm off}\simeq 1.2 {\rm GeV}$, 
off-diagonal gluons $A_\mu^\pm$ can propagate only within 
a short range as $r < M_{\rm off}^{-1} \simeq 0.2{\rm fm}$, 
and becomes {\it infrared inactive} like weak bosons in the Standard Model.
Then, in the MA gauge, off-diagonal gluons $A_\mu^\pm$ cannot contribute to 
the infrared NP-QCD, which leads to infrared abelian dominance.

\subsection{QCD-Monopole Structure in terms of the Off-diagonal Gluon \\
and Infrared Monopole Condensation}

In the MA gauge, there appears a global network of 
monopole world-lines covering the whole system 
as shown in Fig.3(a), and 
this monopole-current system (the monopole part) holds essence of NP-QCD. 
We examine the dual Higgs mechanism by monopole condensation 
in this NP-QCD vacuum in the MA gauge using SU(2) lattice QCD [10,11].
So far, the ``electric sector'' in QCD has been well studied with 
the Wilson loop, 
since QCD is described by the ``electric variable'' such as quarks and gluons. 
To investigate the hidden ``magnetic sector'',
it is useful to introduce the ``dual (magnetic) variable'' 
such as the {\it dual gluon field} $B_\mu $, which is the dual partner 
of the diagonal gluon $A_\mu^3$ and directly couples 
with the magnetic current $k_\mu $.
Owing to the absence of the electric current $j_\mu$ 
in the monopole part,
the dual gluon $B_\mu $ can be introduced 
as the regular field satisfying 
$(\partial \land B)_{\mu\nu}={^*\!F}_{\mu\nu}$ 
and the dual Bianchi identity, 
$
{\partial^{\mu}} {^*\!(}\partial \land B)_{\mu\nu}=j_\nu=0.
$
In the dual Landau gauge $\partial_\mu B^\mu=0$, 
the field equation is simplified as 
$\partial^2 B_\mu ={\partial^\alpha} {^*\!F}_{\alpha \mu}=k_\mu$, 
and the dual gluon field $B_\mu$ can be obtained 
from the monopole current $k_\mu$ as 
\begin{equation}
B_\mu (x) = ( \partial^{-2} k_\mu)(x)= -\frac{1}{4\pi^2} 
\int d^4y \frac{k_\mu(y)}{(x-y)^2}. 
\end{equation}
Here, the mass generation of the dual gluon $B_\mu$ physically means   
the dual Higgs mechanism by monopole condensation, and 
leads to the {\it longitudinal magnetic screening}, 
which can be observed in the following phenomena \cite{CONF2000,SAIT00,SITA98}.
\begin{enumerate}
\item
Due to the longitudinal screening effect on the magnetic flux,  
the inter-monopole potential $V_M(r)$ is screened and 
behaves as a short-range Yukawa potential. 
\item
The Euclidean dual gluon propagator 
$\langle B_\mu(x)B_\nu(y)\rangle_{\rm MA}$ is   
exponentially reduced as in Eq.(6).
\end{enumerate}
Through these tests using lattice QCD, 
we investigate the dual gluon mass $m_B$.
\begin{figure}
\begin{center}
\includegraphics[scale=1]{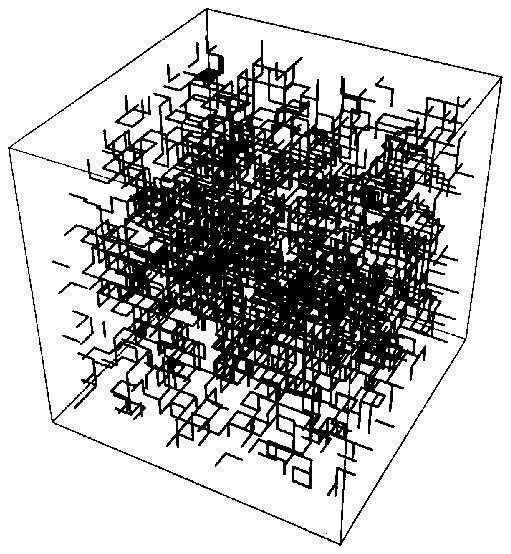}
\includegraphics[scale=0.4]{Fig3b.EPSF}
\includegraphics[scale=0.45]{Fig3c.EPSF}
\includegraphics[scale=0.65]{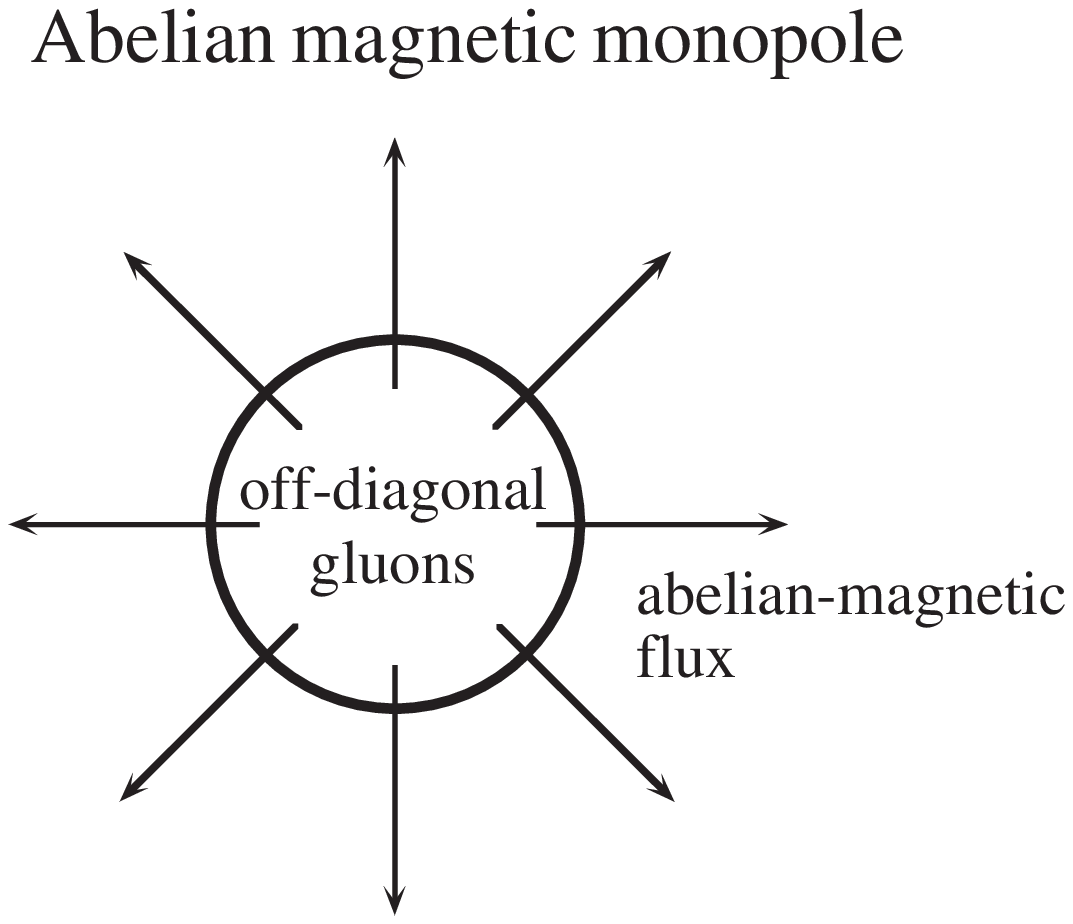}
\caption{The SU(2) lattice-QCD results in the MA gauge. 
(a) The monopole world-line projected into ${\bf R}^3$ 
on the $16^3 \times 4$ lattice with $\beta =2.2$ 
(the confinement phase). 
There appears a global network of monopole currents covering 
the whole system. 
(b) The inter-monopole potential $V_M(r)$ 
v.s. the three-dimensional distance $r$ 
in the monopole-current system on the $20^4$ lattice. 
The solid curve denotes the Yukawa potential with $m_B=0.5$GeV. 
The dotted curve denotes the Yukawa-type potential 
including the monopole-size effect. 
(c) The scalar-type dual gluon correlation 
$\ln (r_E^{3/2} \langle B_\mu(x)B_\mu(y)\rangle_{\rm MA})$ as the 
function of the four-dimensional Euclidean distance $r_E$ on the 
$24^4$ lattice. The slope corresponds to the dual gluon mass $m_B$.
(d) The schematic figure of the QCD-monopole structure in the MA gauge. 
The QCD-monopole includes a large amount of off-diagonal gluons 
around its center as well as the diagonal gluon.}
\end{center}
\end{figure}

First, by putting test magnetic charges in the monopole-current system 
in the MA gauge in SU(2) lattice QCD, 
we define the {\it dual Wilson loop} $W_D(C)$ \cite{SAIT00,SITA98} as  
\begin{equation}
W_D(C) \equiv \exp\{i{e \over 2}\oint_C dx_\mu B^\mu \}=
\exp\{i{e \over 2}\int\!\!\!\int d\sigma_{\mu\nu}{^*\!F}^{\mu\nu}\},
\end{equation}
which is the {\it dual version of the abelian Wilson loop} 
$W_{\rm Abel}(C) 
\equiv \exp\{i{e \over 2}\oint_C dx_\mu A_3^\mu \} 
=\exp\{i{e \over 2}\int\!\!\!\int d\sigma_{\mu\nu}{F}^{\mu\nu}\}$. 
The monopole-antimonopole potential $V_M(r)$ can be derived as 
$
V_{M}(r) = -\lim_{T \rightarrow  \infty} {1 \over T}\ln 
\langle W_D(r,t) \rangle 
$
from the dual Wilson loop. 
In the monopole part in the MA gauge, 
$V_M(r)$ behaves as the Yukawa potential 
$V_{M}(r) \simeq -\frac{(e/2)^2}{4\pi} \frac{e^{-m_Br}}{r}$ 
at the long distance, as shown in Fig.3(b).
From the infrared behavior of $V_M(r)$,  
the dual gluon mass is estimated as $m_B \simeq {\rm 0.5GeV}$ in the MA gauge.

Second, we investigate also the Euclidean scalar-type dual gluon propagator 
$\langle B_\mu(x)B_\mu(y)\rangle_{\rm MA}$ as shown in Fig.3(c), 
and estimate the dual gluon mass as $m_B \simeq 0.5$ GeV 
from its long-distance behavior \cite{SAIT00,SITA98}. 

From these two tests, we obtain the dual gluon mass 
$m_B \simeq 0.5$ GeV in the infrared region, as the direct
evidence of the dual Higgs mechanism by monopole condensation. 

To conclude, lattice QCD in the MA gauge exhibits 
{\it infrared abelian dominance} and 
{\it infrared monopole condensation}, 
and hence the dual Ginzburg-Landau (DGL) theory [14,22-30] 
can be constructed as the infrared effective theory based on QCD.

\begin{figure}
\begin{center}
\includegraphics[scale=0.75]{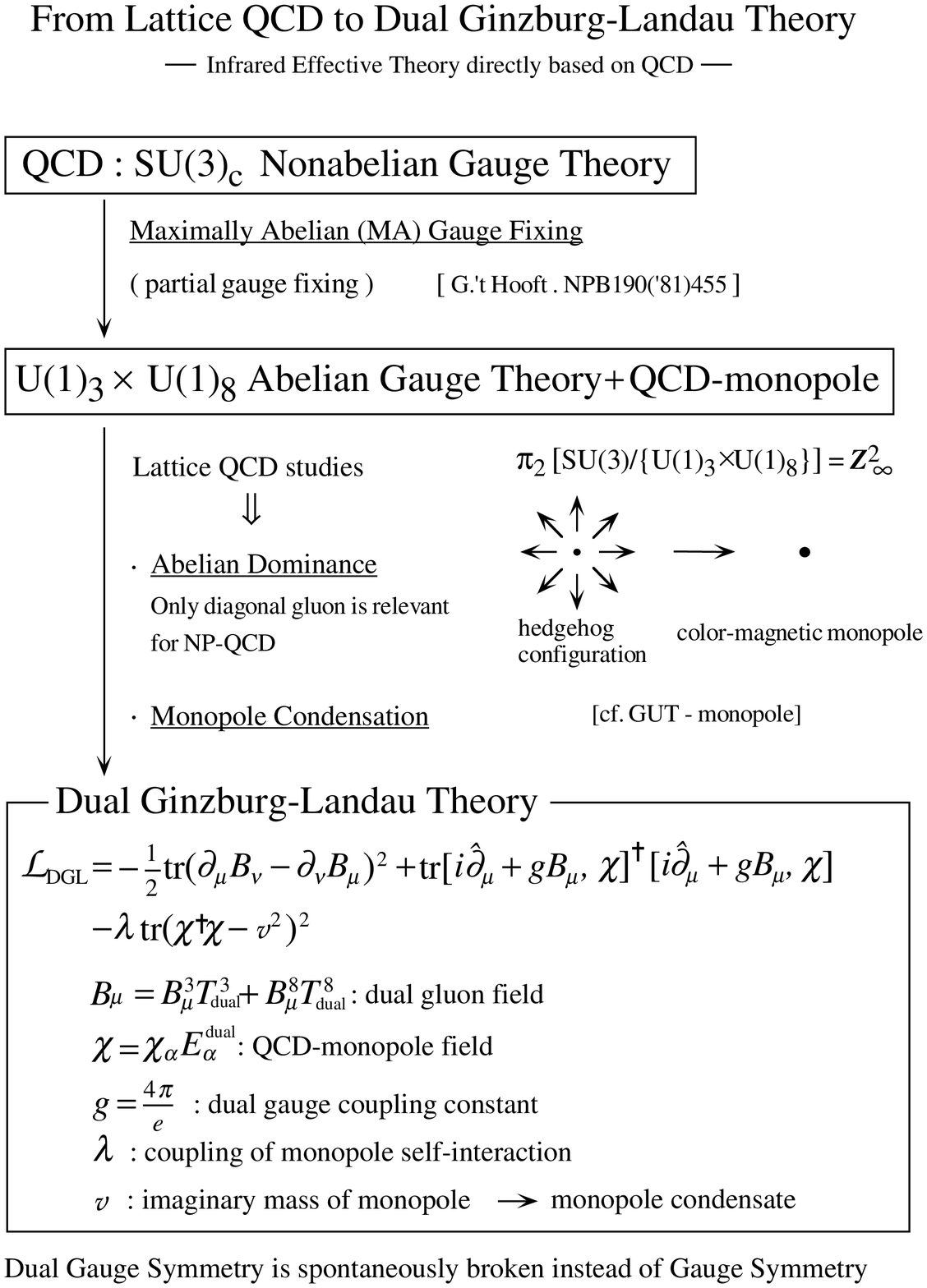}
\caption{The dual Ginzburg-Landau (DGL) theory from lattice QCD in the MA gauge.}
\end{center}
\end{figure}

Here, we compare the QCD-monopole with the point-like Dirac monopole. 
In QED, there is no point-like monopole, 
because the QED action diverges around the point-like monopole. 
{\it The QCD-monopole also accompanies a large abelian action density, 
however, the total QCD action is kept finite even around the QCD-monopole, 
owing to cancellation with the off-diagonal gluon contribution.}
This is the reason why monopoles can appear in QCD.
In [2,9-11], using SU(2) lattice QCD, we investigate
the QCD-monopole structure in the MA gauge 
in terms of the SU(2) action density $S_{\rm SU(2)}$, 
the abelian action density $S_{\rm Abel}$, 
and the off-diagonal contribution 
$S_{\rm off} \equiv S_{\rm SU(2)}-S_{\rm Abel}$. 
We summarize the results on the QCD-monopole structure and 
its related topics as follows.
\begin{enumerate}
\item
Around the QCD-monopole, the abelian action density $S_{\rm Abel}$ 
takes a large value, but the off-diagonal gluon 
contribution $S_{\rm off}$ cancels with the abelian fluctuation 
and keeps the total QCD-action density $S_{\rm SU(2)}$ small.
\item
The QCD-monopole has an {\it intrinsic structure 
relating to the large amount of off-diagonal gluons}  
around its center like the 't~Hooft-Polyakov monopole. (See Fig.3(d).) 
At a large scale, off-diagonal gluons 
inside the QCD-monopole become invisible, 
and QCD-monopoles can be regarded as point-like Dirac monopoles. 
\item
From the concentration of off-diagonal gluons 
around QCD-monopoles in the MA gauge, 
we can predict a  
{\it local correlation between monopoles and instantons}: instantons 
tend to appear around the monopole world-line 
in the MA gauge, because instantons need 
full SU(2) gluon components for existence [9-11,18-21].
\end{enumerate}

\subsection{
Gluonic Higgs and Gauge Invariant Description of MA Projection}

In the MA gauge, the gauge group 
$G \equiv {\rm SU}(N_c)$ is partially fixed into its subgroup 
$H \equiv {\rm U}^{N_c-1}_{\rm local}\times {\rm Weyl}_{N_c}^{\rm global}$, 
and then the gauge invariance becomes unclear.\footnote{ 
In Refs.\cite{CONF2000,IS9900,S969798}, we show a 
useful {\it gauge-invariance criterion} 
on the operator $O_{\rm MA}$: 
{\it If $O_{\rm MA}$ defined in the MA gauge is $H$-invariant, 
$O_{\rm MA}$ is also invariant 
under the whole gauge transformation of $G$.} 
}
In this section, we propose a {\it gauge invariant description 
of the MA projection in QCD}. 
Even without explicit use of gauge fixing, we can define the MA 
projection by introducing a ``gluonic Higgs scalar field'' $\vec \phi(x)$.
For a given gluon field configuration $\{ A_\mu(x) \}$, 
we define a gluonic Higgs scalar 
$\vec \phi(x) \equiv \Omega(x) \vec H \Omega^\dagger(x)$ 
with $\Omega(x) \in {\rm SU}(N_c)$ 
so as to minimize 
\begin{equation}
R[\vec \phi(\cdot)] \equiv \int d^4x \ {\rm tr} 
\left\{[\hat D_\mu, \vec \phi(x)][\hat D_\mu, \vec \phi(x)]^\dagger \right\}. 
\end{equation}
We summarize the features of this description 
as follows \cite{CONF2000,IS9900}.
\begin{enumerate}
\item
The gluonic Higgs scalar $\vec \phi(x)$ does not have amplitude degrees 
of freedom but has only color-direction degrees of freedom, and 
$\vec \phi(x)$ corresponds to 
a ``color-direction'' of the nonabelian gauge connection $\hat D_\mu$ 
averaged over $\mu$ at each $x$. 
\item
Through the projection along $\vec \phi(x)$, 
we can extract the abelian U(1)$^{N_c-1}$ sub-gauge-manifold 
which is most close to the original SU($N_c$) gauge manifold. 
This projection is manifestly gauge invariant, and 
is mathematically equivalent to the ordinary 
MA projection \cite{CONF2000,IS9900}. 
\item
Similar to $\hat D_\mu$, the gluonic Higgs scalar $\vec \phi(x)$ 
obeys the adjoint gauge transformation, 
and $\vec \phi(x)$ is diagonalized in the MA gauge. 
Then, monopoles appear at the hedgehog singularities of $\vec \phi(x)$ 
as shown in Fig.5 \cite{CONF2000,IS9900,SST95a}.
\item
In this description, 
infrared abelian dominance is interpreted as 
infrared relevance of the gluon mode along 
the color-direction $\vec \phi(x)$, and 
QCD seems similar to a nonabelian Higgs theory. 
\end{enumerate}

\begin{figure}
\begin{center}
\includegraphics[scale=0.38]{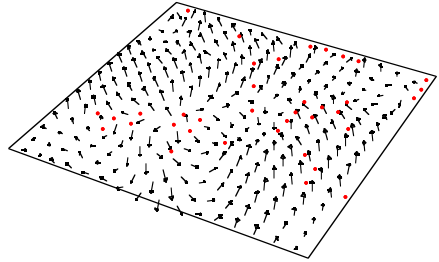}
\caption{
The correlation between the gluonic Higgs scalar field 
$\phi(x)=\phi^a(x)\frac{\tau^a}{2}$ and monopoles (dots)  
in SU(2) lattice QCD with $\beta=2.4$ and $16^4$.
The arrow denotes the SU(2) color direction of 
$(\phi^1(x),\phi^2(x),\phi^3(x))$.
The monopole appears at the hedgehog singularity of 
the gluonic Higgs scalar $\phi(x)$.}
\end{center}
\end{figure}

\section*{Acknowledgments}
We would like to thank Professor Yoichiro~Nambu for his useful suggestions.
We are grateful to Professor Il Tong Cheon for his encouragement.
The lattice calculations have been performed on NEC-SX4
at Osaka University.

\end{document}